\documentclass{aa}

\usepackage[]{graphics}
\usepackage[]{psfig}
\usepackage[]{epsfig}
\usepackage[]{longtable}
\include{epsf}

\begin{document}

\title{Fornax compact object survey FCOS: \\ On the nature of 
 Ultra Compact Dwarf galaxies}

\author {Steffen Mieske \inst{1,2} \and Michael Hilker \inst{1} \and Leopoldo Infante \inst{2}}

\offprints {S.~Mieske}
\mail{smieske@astro.uni-bonn.de}

\institute{
Sternwarte der Universit\"at Bonn, Auf dem H\"ugel 71, 53121 Bonn, Germany
\and
Departamento de Astronom\'\i a y Astrof\'\i sica, P.~Universidad Cat\'olica,
Casilla 104, Santiago 22, Chile
}

\date{Received 21 November 2003 / Accepted 16 January 2004}

\titlerunning{On the nature of Ultra Compact Dwarfs}

\authorrunning{S.~Mieske et al.}

\abstract{The results of the Fornax Compact Object Survey (FCOS) are presented,
complementing the first part of the FCOS described in Mieske et al. (\cite{Mieske02}).
The FCOS aims at investigating the nature of the Ultra Compact Dwarf galaxies (UCDs) 
recently discovered
in the center of the Fornax cluster (Drinkwater et 
al. \cite{Drinkw00a}).
280 unresolved objects with colours $0<(V-I)<1.5$ mag 
in the magnitude space covering UCDs and bright globular clusters 
($18<V<21$ mag) 
were observed spectroscopically. 54 new Fornax members were found, plus
five of the seven already known UCDs.
 Their distribution in radial velocity, colour,
magnitude and space was investigated.
It is found that bright compact objects ($V<20$ or $M_V<-11.4$ mag), including the UCDs,
have a higher mean velocity than faint compact objects
($V>20$ mag) at 96\% confidence.
The mean velocity of the bright compact objects is consistent with that of the dwarf 
galaxy population in Fornax, but inconsistent 
with that of NGC 1399's globular 
cluster system at 93.5\% confidence.
The compact objects
follow a colour magnitude relation with a slope very similar to that of normal dEs,
 but shifted about 
0.2 mag redwards. 
The magnitude distribution of
compact objects shows a fluent transition between UCDs and GCs. 
A slight overpopulation of $8 \pm 4$ objects for $V<20$ mag
is detected with respect to the extrapolation of NGC 1399's GC luminosity function.
The spatial distribution of bright compact objects is in comparison to the faint ones 
more extended 
at 88\% confidence.
All our findings are consistent with the threshing scenario (Bekki et al. \cite{Bekki03}),
suggesting that a substantial fraction of compact
Fornax members brighter than $V\simeq$ 20 mag could be created by threshing dE,Ns. 
Fainter than $V\simeq$ 20 mag, the majority of the objects seem to be genuine GCs. 
Our results are also consistent with
merged stellar super-clusters (Fellhauer \& Kroupa \cite{Fellha02}) 
as an alternative
explanation for the bright compact objects. }

\maketitle


%%%%%%%%%%%%%%%%%%%%%%%%%%%%%%%%%%%%%%%%%%%%%%%%%%%%%%%%%%%%%%%%%%%%%%%%%=
%%
%%%%%%%%%%%%%%%%%%%%%%%%%%%%%%%%%%%%%%%%%%%%%%%%%%%%%%%%%%%%%%%%%%%%%%%%%=
%%
%%%%%%%%%%%%%%%%%%%%%%%%%%%%%%%%%%%%%%%%%%%%%%%%%%%%%%%%%%%%%%%%%%%%%%%%%=
%%
\section{Introduction}
\label{intro}
\subsection{``Normal'' dwarf galaxies}
The faint end of the galaxy luminosity function is mainly populated by 
the dwarf elliptical galaxies (dE) and the dwarf spheroidals (dSph, the faintest dwarf
galaxies in the 
Local Group). They are the most numerous type of galaxies
in the nearby universe, having absolute magnitudes fainter than 
$M_{\rm V}$ $\simeq$ $-$17 mag. They follow a tight magnitude-surface brightness ($M$--$\mu$) relation 
in the sense that central surface brightness increases with increasing total luminosity 
(Ferguson \& Sandage \cite{Fergus88}, \cite{Fergus89}, Drinkwater et al. \cite{Drinkw01a},
Hilker et al.~\cite{Hilker03}). A consequence of this $M$--$\mu$ relation is
that the size of dEs only changes very slowly with total luminosity, with the
effective radius being of the order of a few hundred pc. For photometric studies of 
dEs in galaxy clusters this means that the approximate angular size
of candidate dwarf galaxies is known. This makes the morphological separation between
cluster and back- or foreground galaxies easier.\\
\subsection{Ultra Compact Dwarf galaxies (UCDs)}
However, dwarf galaxies with sizes much smaller than normal dEs at comparable
total luminosities are morphologically misclassified either as stars or distant 
background galaxies when using the $M$--$\mu$ relation as a membership criterion.
In order to test the magnitude of such a bias, among other scientific tasks,
Drinkwater et al. (\cite{Drinkw00a} and \cite{Drinkw00b}) performed the 
all-object Fornax Cluster Spectroscopic Survey
(FCSS) down to $V\simeq 19.5$ ($M_V\simeq -11.9$) mag, using the 2dF 
spectrograph mounted on the Anglo-Australian Telescope.
They were able to confirm the 
existence of a tight magnitude-surface brightness relation for the majority
of the Fornax dwarf
galaxies.\\
In addition they discovered a total of seven unresolved Fornax members, being much smaller
than normal dwarf galaxies at comparable magnitudes (Drinkwater et al. \cite{Drinkw00a}
and \cite{Drinkw03}, Karick et al. \cite{Karick03}). Two of
the UCDs had first been discovered by Hilker et al. (\cite{Hilker99}).
Drinkwater et al. named these objects ``Ultra Compact Dwarf galaxies'' (UCDs). 
They are located in the central
projected 30$'$ of the Fornax cluster around NGC 1399 and cover a luminosity 
range of $18<V<19.5$ mag, or $-13.4<M_V<-11.9$ mag ($(m-M)$=31.4 mag, Ferrarrese et al.
\cite{Ferrar00}). 
This makes them brighter than 
the brightest
 Milky Way globular cluster $\omega$ Centauri ($M_V\simeq -10.2$ mag), but by far 
fainter than the compact dwarf
elliptical galaxy M32 ($M_V\simeq -16$ mag, see Graham \cite{Graham02}). The UCDs
do fall into the luminosity range of 
dwarf galaxy nuclei (Lotz et al. \cite{Lotz01}).\\
\subsection{Possible origins of UCDs}
Three different origins for the UCDs are in discussion:
(1) they are 
the brightest globular clusters (GCs) of NGC 1399's very rich globular cluster system (GCS) (e.g.
Mieske et al. \cite{Mieske02}, Dirsch et al. \cite{Dirsch03}); (2) they are
the remnant nuclei of stripped dwarf galaxies which have lost their outer
parts in the course of tidal interaction with the Fornax cluster's potential (Bekki et al.
\cite{Bekki03}); (3) they 
are formed from the amalgamation of stellar super-clusters that were created in collisions between
gas-rich galaxies (Fellhauer \& Kroupa \cite {Fellha02}, Kroupa \cite{Kroupa98}, 
Maraston et al. \cite{Marast04}).\\
Regarding possibility (1): NGC 1399's GCS contains about 6500 GCs 
(about 60 times more than the Milky Way) and 
extends out to about 30 $'$ (Dirsch et al. \cite{Dirsch03}). The existence of 
genuine GCs in the same projected area and the same magnitude range as the UCDs cannot be 
ruled out a priori.\\
Regarding possibility (2): Bekki et al. (\cite{Bekki01} and \cite{Bekki03}) show that 
threshing of ``normal'' dE,Ns
 in the Fornax cluster's central potential can create compact
remnants with properties like the UCDs, given that the following requirements
are fulfilled. The progenitor dE,N needs to
have a sufficiently
high mass to partially survive the tidal disruption and a shallow dark matter
core in order to allow substantial tidal disruption. In addition, its orbit needs to be
very eccentric and its pericenter smaller compared to average cluster dEs.
Therefore, the spatial
distribution of the UCDs should be more concentrated towards the cluster center
than that of the normal cluster dEs, which is qualitatively consistent
with the radial distribution of the UCDs found by Drinkwater et al.
Based on the simulations of Bekki et al. (\cite{Bekki03} and private communications) 
this means that also the peculiar velocities of the progenitor dE,Ns 
with respect to the cluster's center of mass
velocity would be limited to a smaller range than for average dEs. 
Bekki et al. conclude that due to these
requirements, in the Fornax cluster
the total number of UCDs
should not be much higher than ten, covering a luminosity range of $-15<M_V<-8$ mag.\\
Regarding possibility (3): Fellhauer \& Kroupa (\cite{Fellha02}) 
have shown that in violent star forming regions, like
dense knots in tidal arms of merger events, 
a conglomerate of young massive stellar clusters will rapidly merge to form a young stellar 
super cluster, which itself will turn into an object with similar properties
as the UCDs after aging several Gyrs. Fellhauer \& Kroupa find that even for highly
eccentric orbits, these compact star clusters maintain most of their mass after 10 Gyrs.
Examples for young very massive conglomerates
of stellar clusters are the star formation knots in the Antenna galaxies (Whitmore et al.
\cite{Whitmo99}) and the star cluster W3 in NGC 7252 (Maraston et al. \cite{Marast04}). 
One possible origin for the UCDs
could then be merger events several Gyrs ago in the Fornax cluster which led to the 
creation of massive super-clusters. 
\subsection{High resolution follow up investigations of UCDs}
Follow-up high resolution photometry with the HST-STIS and high resolution UVES spectroscopy
 (Drinkwater et al. \cite{Drinkw03})
showed that the UCDs have effective radii between
10 and 22 pc and internal velocity dispersions between
24 and 37 km/s, placing them at the extreme end of what is found for globular
clusters (Djorgovski et al. \cite{Djorgo97}, Meylan et al. \cite{Meylan94} and \cite{Meylan01}). Their M/L$_V$ ratios are between 2 and 4,
in the same range as derived for the very luminous stellar cluster Mayall II 
in M31 by Meylan et al. (\cite{Meylan01}) and for $\omega$ Centauri (Meylan \cite{Meylan87},
Meylan et al. \cite{Meylan94} and \cite{Meylan95}),
but higher than for the vast majority of GCs, which have values of M/L$\simeq$ 1. Note that
both Mayall II and $\omega$ Centauri are very bright stellar clusters, and are in general not even
considered as normal GCs due to their complex star formation history 
(Hilker \& Richtler \cite{Hilker00}).
The simulations of Bekki et al. predict M/L ratios for the threshed nuclei
in the same range as observationally found for the UCDs.\\
Drinkwater
et al. (\cite{Drinkw03}) find that in the velocity dispersion-absolute magnitude ($\sigma - M_V$) plane, 
the UCDs lie
off  the extrapolation of the relation defined by the GCs and rather appear to
follow the Faber-Jackson relation for elliptical galaxies (Faber \& Jackson \cite{Faber76}).
They populate
the same region as dE-nuclei. However, as this disagreement
of the UCD values is only with respect to the {\it extrapolation} of the GC relation defined at fainter
magnitudes, it is not a 
sufficient condition to separate UCDs from GCs. This would require $\sigma$ measurements of bona fide 
GCs in the
brightness range of UCDs.\\
In conclusion, if the UCDs were bright genuine GCs, they would lie at the 
extreme end of the mass-light-size space of their kind (Drinkwater et al. \cite{Drinkw03}), while if they
were nuclei of stripped dE,Ns (Bekki et al. \cite{Bekki03}) or merged stellar super-clusters (Fellhauer \&
Kroupa \cite{Fellha02}), 
they would be more average examples of their kind.\\
\subsection{The first part of the ``Fornax Compact Object Survey'' FCOS}
Although the results obtained from high-resolution spectroscopy and photometry suggest
that the UCDs are more complicated objects than normal GCs, no definite discriminating
judgment on their nature has been made until now. Therefore, investigating
group properties of bright compact Fornax members like their distribution in radial velocity, 
magnitude, colour and space 
can prove to be a very valuable contribution to the ongoing discussion\footnote{With compact, we here
refer to sources unresolved
on ground based imaging, corresponding to an upper limit of a few tens of parsecs in radius
at the Fornax cluster distance.}.
It would be very interesting to know whether the compact Fornax members are divided
up into subgroups with respect to any of these four global properties.
Do the fainter ones behave more like GCs, while the brighter ones (including the UCDs) 
are different?\\
As Drinkwater et al. (\cite{Drinkw00a} and \cite{Drinkw00b}) were 
not able to close the magnitude gap towards the GCs 
due to the design of their survey, 
the approach of our investigation is to look for 
compact Fornax members in the magnitude regime between
the UCDs and bright GCs, including both groups of objects.
To this end, we have conducted the Fornax Compact Object Survey (FCOS),
whose first part (FCOS-1) was described in Mieske et al. (\cite{Mieske02},
hereafter {\it paper I}).
In FCOS-1, about 20\% of compact Fornax candidates within 20$'$ of NGC 1399
were observed spectroscopically. 
12 new compact Fornax members in the magnitude regime $19.7<V<21$ mag were found.
The resulting luminosity distribution, shown in Fig.~5 of {\it paper I},
 supported a fluent transition between UCDs and GCs, speaking against 
a magnitude gap between both populations. 
However, the number counts involved were too low to give a more quantitative
statement. With only 12 objects at hand, 
also the velocity, colour and spatial distribution
of the compact Fornax members was too poorly sampled to
investigate possible segregations into subgroups.\\
\subsection{Aim of this paper}
In this paper we present FCOS-2, the second part of our survey, and merge
the datasets of FCOS-1 and FCOS-2. FCOS-2 increases the area 
coverage of FCOS-1 by a factor of about three. There are four main aspects that are
 addressed with the entire FCOS sample at hand:\\
1. Investigate the radial velocity distribution in order to to look for dynamically
distinct subgroups.\\
2. Determine the luminosity distribution of compact Fornax members in order
to check whether a significant number of stripped dE-nuclei mixes up 
with the bright GCs.\\
3. Investigate the colour-magnitude (CM) relation of compact Fornax members in order
to compare it both with the colours of GCs and dwarf galaxies.\\
4. Investigate the radial distribution of compact Fornax members in order to look
for differences in the spatial distribution between subgroups.\\
The paper is structured as follows: in the next section, the candidate selection
and the observations for FCOS-2 are described. In Sect.~\ref{datared}, the data reduction
is described. The results of the entire FCOS are shown in Sect.~\ref{results}, based
on the merging of the FCOS-1 and FCOS-2 database. The results are discussed in 
Sect.~\ref{discussion}. We finish the paper with summary and conclusions in 
Sect.~\ref{summary}.\\
\section{Candidate selection and observations}
\label{candsel}
To select candidates for FCOS-2, we
used two different image sets with the aim to optimize
our spatial coverage.\\
First, the same 25$'$ $\times$ 25$'$
wide field images
used in FCOS-1, obtained in Johnson $V$ and $I$ with
the 2.5m Du Pont Telescope at Las Campanas Observatory, Chile, in
December 1999 (see {\it paperI} and Hilker et al. \cite{Hilker04} for more details).\\
Second, photometric data from Dirsch et al. 
(\cite{Dirsch03}, and private communications) obtained in the
Washington $C$ and Kron $R$ filters
with the 36$'$ $\times$ 36$'$ MOSAIC camera at the 4m CTIO telescope.\\
The CTIO-data cover the entire central region of NGC 1399,
allowing the selection of candidates very close to its center. This
was not possible with the LCO-data as they miss out much of the 
bright central region of NGC 1399.\\
In Fig.\ref{mapcand},
the location of the LCO- and CTIO-fields is indicated.\\
\subsection{Selection of candidates from the LCO photometry}
\label{lcocand}
In contrast to {\it paper I}, we now selected only {\it unresolved} objects,
defined by the SExtractor star classifier (Bertin \& Arnouts \cite{Bertin96}) 
being larger than 0.45 either in
$V$ or in $I$. This
was done because in {\it paper I} 100\%  of the extended objects investigated
spectroscopically turned out
to be background galaxies far behind the Fornax cluster. 
Of the unresolved objects, only those were selected which had not been observed
either in the FCSS nor in the FCOS before and which satisfy the
conditions $V<21$ mag and $(V-I)<1.40$. 
The magnitude limit is identical to {\it paper I}, being
about 3 mag brighter than the GCLF turn-over and about 1.5 mag fainter
than the faintest UCD.
The red colour limit is 0.10 mag bluer than the 1.50 
mag adopted in {\it paper I} in order to decrease contamination by non-cluster
members, as the reddest newly discovered 
Fornax member from {\it paper I} had $(V-I)=1.27$ mag. \\
\subsection{Selection of candidates from the CTIO photometry}
\label{ctiocand}
As for the LCO-images, only objects unresolved on the CTIO-data 
(Dirsch et al. \cite{Dirsch03}) were regarded as possible candidates. 
The CTIO-data had been obtained in the Washington $C$ and Cron $R$ filters,
 in contrast to our Johnson $V$ and $I$ data from LCO.
The magnitude limit of $V<21$ mag was transformed into 
$R<20.5$ mag, with $(V-R)=0.5$ mag taken from
Worthey (\cite{Worthe94}) for a typical GC population of 12 Gyrs and [Fe/H]=$-$1 dex.
The colour window was adopted as $1.0<(C-R)<2.0$, 
based on the results of Dirsch et al. (\cite{Dirsch03}). This range corresponds
approximately to $0.8<(V-I)<1.3$ mag (Dirsch, private communications), extending to
about 0.1 mag redder than the colour range of UCDs, see Table~1 of {\it paper I}
or Fig.~\ref{cmd} of this paper. Therefore, it is unlikely that any UCD is missed by applying 
this selection window.\\
It holds that $(C-R) - (V-I)=0.5$ mag, according to Worthey (\cite{Worthe94}),
for the same typical GC population with $(V-R)=0.5$.\\\\
\begin{figure}[h!]
\psfig{figure=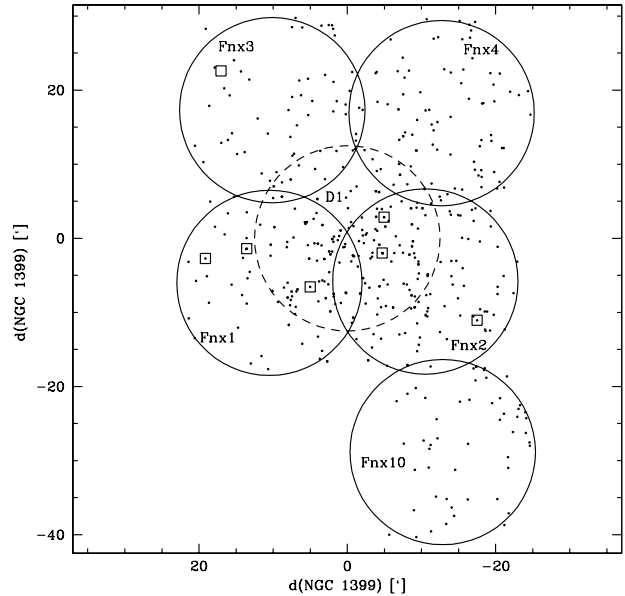,width=8.7cm}
\caption[]{\label{mapcand}Map of the central Fornax cluster, with coordinates centered on NGC 1399.
Dots indicate the positions
of all previously unobserved candidates satisfying our photometric selection criteria 
for possible globular
clusters.
Squares indicate the position of the 7 UCDs. 
Solid circles indicate the approximate
field limits of the LCO-images and mask regions, which in reality are not perfectly circular.
The dashed circle indicates the field where additional candidates were chosen from the
CTIO-data (Dirsch et al. \cite{Dirsch03}). The UCD in field Fnx3 (UCD 5)
was not classified as
unresolved by SExtractor, being given a star-classifier value of 0.18 in $V$ and 0.23 in $I$. 
Note that also UCD 3 in Fnx1 is given low star classifier
values of 0.46 in $V$ and 0.36 in $I$, just above the minimum value of 0.45 required to
satisfy our selection criteria.}
\end{figure}
\noindent In total, 462 unresolved candidate objects in the central Fornax cluster 
were selected for the FCOS, covering a colour-magnitude
range of $0<(V-I)<1.5$ and $18<V<21$ mag.
\subsection{Observations}
The observations for FCOS-2
were performed in the three nights of 2002/12/04 to 2002/12/06 
at the 2.5m Du Pont telescope at Las Campanas. The
instrument was the Wide Field CCD (WFCCD) camera, which reimages a
25$'$ field onto a Tek$\#$5-Detector of 2048x2048 pixel with a
scale of 0.774$''$/ pixel. Multi-slit masks and the ``blue'' grism,
which has a
transmission between 3600 and 7800 {\AA} and a
dispersion of about 3 {\AA} per pixel, were used. As the slit width
in our observation was $\approx$ 1.5$''$ (= 2 pixel), the effective
resolution was of the order of about 6 \AA, corresponding to 450 km/s
at 4000 \AA.\\
We did not use the higher resolution ``H\&K'' grism used in FCOS-1, 
as the spectral resolution 
achieved in FCOS-1 was not sufficient to study possible line index differences between
UCDs and bright GCs ({\it paperI}). The ``blue'' grism
with its lower resolution and higher wavelength coverage allowed smaller
integration times.\\
The locations of the candidates for each field are indicated in 
Fig.~\ref{mapcand}. There is some overlap between the central field
D1 and the adjacent fields Fnx1 to Fnx4. Measurements of the
same object in different fields are used in Sect.~\ref{syserr} for estimating 
the systematic
uncertainties in radial velocity determination. A slit could not be allocated on the mask 
for all candidates (see Fig.~\ref{mapobs1}), as a minimum slit-length of 9 arcsec and
a minimum distance between
neighbouring candidates of
3 arcsec perpendicular to the dispersion direction was demanded.\\
A total of six fields as indicated in Fig.~\ref{mapcand} 
was observed in the three nights, 
using eight different slit masks (three masks for D1, the field with the highest
candidate density).
The total integration time per mask was 2 hours. 
The spectra were calibrated with lamp exposures taken directly before or
after the science exposure.
The same
radial velocity standards as in FCOS-1 were observed in the course of the three nights.\\
\label{data}
\section{Data reduction}
\label{datared}
Data reduction was performed with the IRAF-packages IMRED, TWODSPEC 
and ONEDSPEC. 
The mask-exposures were bias corrected and combined using a cosmic-ray 
removal 
algorithm. The 
combined images were response calibrated using the dome-flat 
spectra averaged perpendicularly to the dispersion direction. Then, each single
2D science spectrum including the corresponding sky region and the calibration lamp exposures
were cut out of the entire image. 
The spectra were extracted using the task APALL in
the ONEDSPEC package and then wavelength calibrated with the
task IDENTIFY in the ONEDSPEC package.\\
The radial velocity of the observed objects was determined with cross-correlation,
applying the IRAF-task FXCOR in the RV-package. The same three objects as in FCOS-1
were used as radial
velocity templates for the cross-correlation: the re-observed spectra of 
HD 54810 and HD 22879, and a synthetic spectrum taken from Quintana et al. 
(\cite{Quinta96}). The mean of the three cross correlations, corrected for
the respective heliocentric velocities of the templates, was adopted as the
final radial velocity.\\
\subsection{Systematic radial velocity errors}
\label{syserr}
\subsubsection{Errors from double observations within FCOS}
\label{doubleerror}
The overlap between the central field D1 and the adjacent fields resulted 
in a total of ten objects
that were observed spectroscopically twice. 
These double measurements allow 
an estimate of the systematic uncertainty of the radial velocity results.
 See top panel of Fig.~\ref{comprvucb_oro} for the corresponding plot. 
The mean single measurement error is 90 km/s. 
The median difference between D1 from FCOS-2
and FCOS-1 is 85 $\pm$ 34 km/s, while the two objects observed twice in FCOS-2
agree nicely. This shows that there is good consistency between the masks
used in FCOS-2, but that there is a systematic difference between FCOS-1 and 2
of the order of the single measurement error.
A difference
of 100 km/s corresponds to a shift of about 1.5 \AA, or half a pixel with
the blue grism and about one pixel with the H\&K grism. The slit width is
 about 2 pixel in FCOS-1 and FCOS-2. Thus, the order of this
systematic difference is about 25-50\% of the slit-width and therefore
within the expectable error range, possibly caused by instrument flexion.\\
\subsubsection{Errors from comparison with radial velocities from Dirsch et al.}
\begin{figure}[]
\psfig{figure=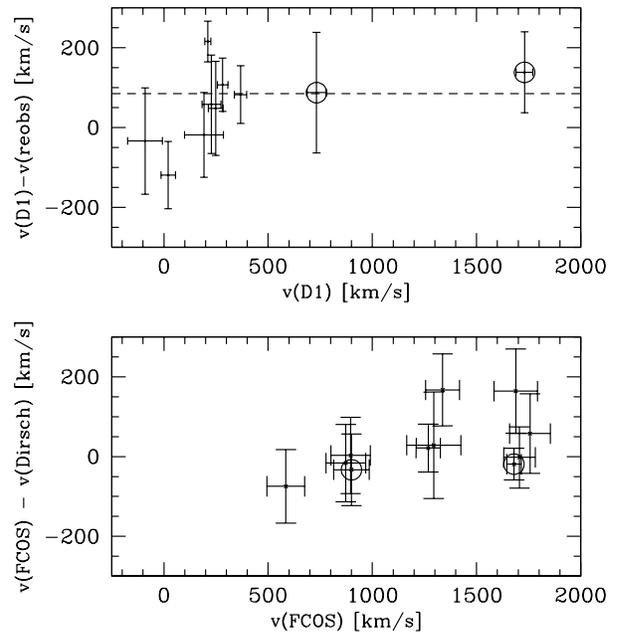,width=8.7cm}
\caption[]{\label{comprvucb_oro}{\it Top panel:} Comparison between the two radial velocity measurements
obtained in the FCOS for the objects that were observed twice due to the
overlap of the central field D1 with adjacent fields. 
The x-axis shows the radial velocity obtained in field D1, during FCOS-2. The y-axis shows
the
difference between the D1-velocity and the velocity obtained in the adjacent field.
The objects marked with circles were observed first in FCOS-1,
the remaining two in FCOS-2. The dashed line indicates the median difference between
D1- and FCOS-1-velocities. {\it Bottom panel:} Comparison between the FCOS radial velocities and the
radial velocities from Dirsch et al. (private communication) for the objects
observed in common. The two objects marked with circles are from FCOS-1,
the rest from FCOS-2.}
\end{figure}

As an additional test,
we compare our radial velocities with those observed by Dirsch et al. (\cite{Dirsch04}, private
communications),
who have obtained about 500 spectra
of GCs around NGC 1399 brighter than $V\simeq$ 23 mag, with 20 GCs in the same
magnitude regime as the FCOS. They used FORS2/MXU at the VLT. 
See bottom panel of Fig.~\ref{comprvucb_oro} for the corresponding plot. 
The mean difference between our and their radial velocity measurements for
the eleven objects observed in common is 27 $\pm$ 23 km/s, almost consistent with zero.
The scatter about the mean difference is 77 km/s, 
the mean single measurement error of our results is 86 km/s. 
This shows again that
the errors resulting from our cross-correlation are a very good estimate
of the real radial velocity uncertainty. Fitting a linear
relation to the observed differences yields a slope different from zero at only the 0.45
$\sigma$ significance, indicating that velocity difference is not correlated with
velocity within our velocity errors. If one excludes the two FCOS-1 objects
from the analysis, taking only FCOS-2 measurements, the mean radial velocity difference
becomes 37 $\pm$ 27 km/s.\\
In conclusion, the radial velocities measured in FCOS-2 are about 85 km/s higher than in FCOS-1
and about 30-40 km/s higher than those from Dirsch et al..
In other words, the values from Dirsch lie between the 
values from FCOS-1 and FCOS-2.\\
As a consequence, we adopt the mean of the FCOS-1 and FCOS-2 velocity for the two 
Fornax members observed twice in the course of the FCOS.\\
\section{Results}
\label{results}
\begin{figure}[]
\psfig{figure=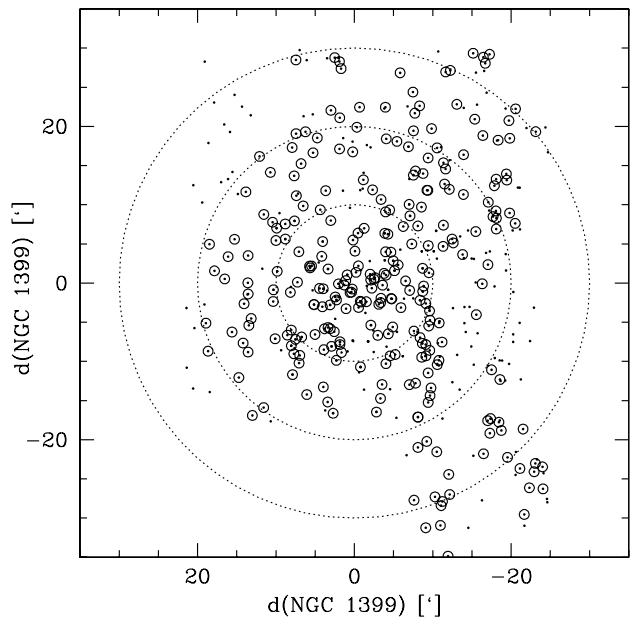,width=8.7cm}
\caption[]{\label{mapobs1}Map of the central square degree around NGC 1399. 
Dots are all candidates
for spectroscopic investigation that satisfy the selection criteria for FCOS as 
defined in sections~\ref{lcocand} and~\ref{ctiocand}. Small circles indicate the objects
observed in the FCOS. The large dotted circles indicate distances of 10, 20 and
30 arcminutes from NGC 1399.}
\end{figure}

\begin{figure}[]
\psfig{figure=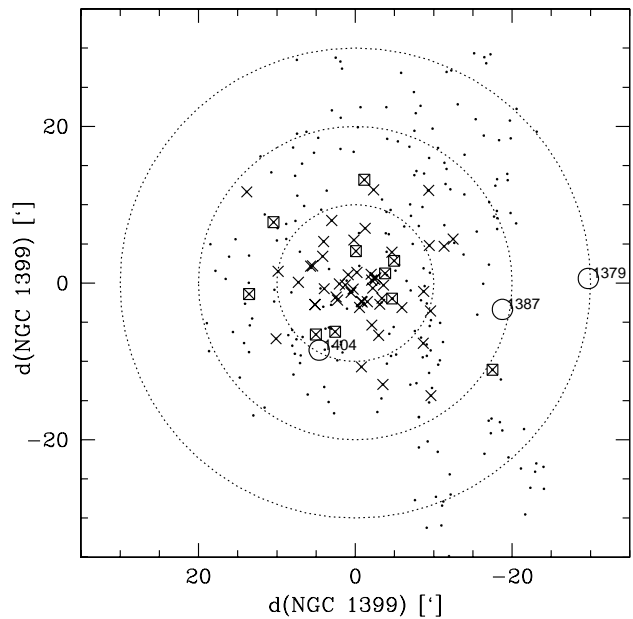,width=8.7cm}
\caption[]{\label{mapobs2}Map of the same region as in Fig.~\ref{mapobs1}. Dots
are all objects observed in the FCOS, marked with small circles in the previous Fig.~\ref{mapobs1}.
Crosses indicate the observed objects with radial velocities in the Fornax cluster 
range, i.e. between 550 and 2400 km/s. Squares mark the Fornax members brighter than
$V=20$ mag. Circles mark the location of Fornax giant galaxies with their NGC numbers
indicated except for NGC 1399, which is located in the origin.}
\end{figure}

In the following subsections, the results of FCOS are presented,
merging the smaller FCOS-1 data set with the larger one from FCOS-2.
The total number of objects with successfully determined radial velocity in the FCOS
is 280 -- out of 462 candidates --, covering a colour-magnitude range of $0<(V-I)<1.5$ and $18<V<21$ mag.
To separate Fornax members from Milky Way stars and background galaxies, we applied a lower limit of
550 km/s and an upper limit of 2400 km/s in radial velocity for Fornax membership (see Fig.~\ref{histvrad}), 
resulting in 59 Fornax members out of the 280 observed objects.
In Table~\ref{fornmem}, the 59 Fornax cluster 
members are tabulated. They have colours $0.8<(V-I)<1.4$, consisting of 12 members
from FCOS-1, 42 new members from FCOS-2 and five of the seven already known UCDs. A list
of all investigated objects in FCOS-2, including foreground stars and background galaxies,
will be added to the existing list of FCOS-1 objects at the CDS.\\
A map overplotting all candidates and observed candidates is 
given in Fig.~\ref{mapobs1}. A map showing only the observed objects plus 
the Fornax members marked is given in Fig.~\ref{mapobs2}.\\
%%%% next line to be removed in 2-column version, to be included in astro-ph version with 0.9 or 0.95
\renewcommand{\baselinestretch}{0.92}
\begin{table*}[ht!]
\caption{\label{fornmem}All Fornax cluster members detected in the FCOS,
ordered by magnitude. The
object codes are ``FCOS x-xxx'' for FCOS1 (Mieske et al.\cite{Mieske02}) and ``FCOS x-xxxx'' for 
FCOS2 (this paper). Errors in $V$ and $(V-I)$ are between 0.04 and 0.05 mag, as judged
from the artificial star experiments described in Sect.~\ref{surveycompleteness}.
$^*$For all objects FCOS 0-2xxx, $V$ and $(V-I)$ are
derived indirectly from the $C-R$ photometry of Dirsch et al. (\cite{Dirsch03}), using
$V=R+0.5$ mag and $(V-I)=(C-R)-0.5$ mag (see text for further details).
$^{**}$$v_{\rm rad}$ is the mean of the values measured in FCOS-1 and FCOS-2 
(see text for further details).}
\begin{tabular}[l]{lllrrrrl}
Name&$\alpha$ (2000.0)&$\delta$ (2000.0)&$v_{\rm rad}$&$\Delta v_{\rm rad}$&$V$$^*$&$(V-I)$$^*$&Comments\\\hline\hline
FCOS 1-2053 & 03 : 38 : 54.05 & -35 : 33 : 33.8 & 1493 &   61 & 18.06 & 1.20 & UCD 3\\ 
FCOS 2-2143 & 03 : 38 : 05.04 & -35 : 24 : 09.7 & 1244 &   55 & 18.93 & 1.16 & UCD 6\\ 
FCOS 1-2083 & 03 : 39 : 35.90 & -35 : 28 : 24.2 & 1848 &   85 & 19.12 & 1.14 & UCD 4\\ 
FCOS 2-2111 & 03 : 38 : 06.29 & -35 : 28 : 58.8 & 1280 &   58 & 19.23 & 1.14 & UCD 2\\ 
FCOS 2-2031 & 03 : 37 : 03.24 & -35 : 38 : 04.6 & 1571 &   75 & 19.31 & 1.17 & UCD 1\\ 
FCOS 1-021 & 03 : 38 : 41.96 & -35 : 33 : 12.9 & 2010 &   40 & 19.70 & 1.18 &  \\ 
FCOS 3-2004 & 03 : 39 : 20.50 & -35 : 19 : 14.2 & 1341 &   57 & 19.73 & 1.23 &  \\ 
FCOS 3-2027 & 03 : 38 : 23.74 & -35 : 13 : 49.4 & 1553 &   72 & 19.76 & 1.04 &  \\ 
FCOS 0-2031 & 03 : 38 : 28.97 & -35 : 22 : 55.9 & 1654 &   69 & 19.87 & 1.13 &  \\ 
FCOS 2-2134 & 03 : 38 : 10.73 & -35 : 25 : 46.2 & 1683 &   86 & 19.90 & 1.34 &  \\ 
FCOS 0-2030 & 03 : 38 : 28.34 & -35 : 25 : 38.3 & 1771 &   56 & 20.01 & 1.10 &  \\ 
FCOS 2-2153 & 03 : 38 : 06.48 & -35 : 23 : 03.8 & 1390 &  124 & 20.04 & 1.08 &  \\
FCOS 1-2024 & 03 : 38 : 25.54 & -35 : 37 : 42.6 & 1789 &  124 & 20.06 & 1.20 & \\
FCOS 0-2066 & 03 : 38 : 23.23 & -35 : 20 : 00.6 & 1255 &   69 & 20.09 & 1.12 &  \\ 
FCOS 1-060 & 03 : 39 : 17.66 & -35 : 25 : 30.0 &  980 &   45 & 20.19 & 1.27 &  \\ 
FCOS 0-2033 & 03 : 38 : 30.72 & -35 : 27 : 46.1 & 1400 &  113 & 20.28 & 1.29 &  \\ 
FCOS 1-063$^{**}$ & 03 : 38 : 56.14 & -35 : 24 : 49.1 &  688 &   45 & 20.29 & 1.05 &  \\ 
FCOS 0-2024 & 03 : 38 : 16.51 & -35 : 26 : 19.3 &  902 &   93 & 20.32 & 0.74 &  \\ 
FCOS 2-073 & 03 : 38 : 11.98 & -35 : 39 : 56.9 & 1300 &   45 & 20.40 & 1.20 &  \\ 
FCOS 4-2028 & 03 : 37 : 43.49 & -35 : 15 : 10.1 & 1240 &   89 & 20.41 & 1.18 &  \\ 
FCOS 3-2019 & 03 : 39 : 37.18 & -35 : 15 : 22.0 & 1921 &  114 & 20.49 & 1.10 &  \\ 
FCOS 2-2106 & 03 : 38 : 25.06 & -35 : 29 : 25.1 & 1313 &  168 & 20.51 & 0.99 &  \\
FCOS 2-2107 & 03 : 38 : 25.66 & -35 : 29 : 19.7 & 1267 &  101 & 20.51 & 1.32 &  \\
FCOS 0-2007 & 03 : 38 : 54.67 & -35 : 29 : 44.2 & 1761 &   98 & 20.54 & 0.91 &  \\ 
FCOS 2-2161 & 03 : 37 : 33.86 & -35 : 22 : 19.2 & 2009 &  106 & 20.54 & 1.02 &  \\ 
FCOS 2-2165 & 03 : 37 : 28.22 & -35 : 21 : 23.0 & 1356 &  139 & 20.57 & 0.97 &  \\ 
FCOS 1-019 & 03 : 38 : 54.59 & -35 : 29 : 45.8 & 1680 &   35 & 20.62 & 1.01 &  \\ 
FCOS 1-2095 & 03 : 38 : 33.82 & -35 : 25 : 57.0 & 1245 &  220 & 20.66 & 1.09 & \\
FCOS 0-2023 & 03 : 38 : 12.70 & -35 : 28 : 57.0 & 1705 &   76 & 20.66 & 1.11 &  \\ 
FCOS 0-2062 & 03 : 38 : 18.43 & -35 : 27 : 39.6 & 1338 &   81 & 20.67 & 0.89 &  \\ 
FCOS 1-058$^{**}$ & 03 : 38 : 39.30 & -35 : 27 : 06.4 & 1661 &   60 & 20.67 & 1.04 &  \\ 
FCOS 0-2069 & 03 : 38 : 26.71 & -35 : 30 : 07.2 & 1914 &   91 & 20.67 & 0.99 &  \\ 
FCOS 2-078 & 03 : 37 : 41.83 & -35 : 41 : 22.2 & 1025 &   60 & 20.69 & 1.21 &  \\ 
FCOS 2-2094 & 03 : 37 : 42.24 & -35 : 30 : 33.8 & 1462 &  121 & 20.71 & 1.16 &  \\ 
FCOS 0-2072 & 03 : 38 : 32.06 & -35 : 28 : 12.7 & 1559 &  102 & 20.71 & 1.16 &  \\ 
FCOS 2-2072 & 03 : 38 : 14.69 & -35 : 33 : 40.7 & 1331 &  149 & 20.72 & 0.91 &  \\ 
FCOS 0-2041 & 03 : 38 : 44.11 & -35 : 19 : 01.6 & 1287 &   66 & 20.73 & 1.32 &  \\ 
FCOS 1-2103 & 03 : 38 : 57.38 & -35 : 24 : 50.8 &  896 &   94 & 20.74 & 1.18 &  \\ 
FCOS 0-2063 & 03 : 38 : 19.08 & -35 : 26 : 37.3 & 1692 &  104 & 20.80 & 1.00 &  \\ 
FCOS 0-2032 & 03 : 38 : 30.22 & -35 : 21 : 31.0 & 1402 &   73 & 20.80 & 0.86 &  \\ 
FCOS 2-086 & 03 : 37 : 46.77 & -35 : 34 : 41.7 & 1400 &   50 & 20.81 & 0.92 &  \\ 
FCOS 0-2027 & 03 : 38 : 19.49 & -35 : 25 : 52.3 & 1289 &  130 & 20.83 & 1.43 &  \\ 
FCOS 0-2089 & 03 : 38 : 17.09 & -35 : 26 : 30.8 & 1294 &   95 & 20.83 & 1.16 &  \\ 
FCOS 2-2127 & 03 : 38 : 11.69 & -35 : 27 : 16.2 & 1476 &  201 & 20.83 & 1.18 &  \\ 
FCOS 1-2077 & 03 : 38 : 40.56 & -35 : 29 : 10.0 &  586 &   91 & 20.84 & 1.16 &  \\ 
FCOS 1-2115 & 03 : 38 : 49.18 & -35 : 21 : 42.1 &  872 &   95 & 20.85 & 1.22 & \\ 
FCOS 4-049 & 03 : 37 : 43.09 & -35 : 22 : 12.9 & 1330 &   50 & 20.85 & 0.98 &  \\ 
FCOS 2-089 & 03 : 38 : 14.02 & -35 : 29 : 43.0 & 1235 &   45 & 20.87 & 1.08 &  \\ 
FCOS 0-2025 & 03 : 38 : 17.98 & -35 : 15 : 06.1 & 1401 &  264 & 20.87 & 1.07 &  \\ 
FCOS 0-2026 & 03 : 38 : 19.03 & -35 : 32 : 22.2 & 1726 &   97 & 20.91 & 1.11 &  \\ 
FCOS 0-2074 & 03 : 38 : 35.66 & -35 : 27 : 15.5 & 2274 &  112 & 20.91 & 1.07 &  \\ 
FCOS 1-2089 & 03 : 38 : 48.86 & -35 : 27 : 43.9 & 1559 &  158 & 20.92 & 1.02 &  \\
FCOS 2-2100 & 03 : 38 : 00.17 & -35 : 30 : 08.3 &  997 &  152 & 20.94 & 1.07 &  \\ 
FCOS 2-095 & 03 : 37 : 46.55 & -35 : 28 : 04.8 & 1495 &   45 & 20.96 & 1.14 &  \\ 
FCOS 1-2080 & 03 : 38 : 41.35 & -35 : 28 : 46.6 & 1647 &  124 & 20.96 & 0.99 &  \\ 
FCOS 1-064 & 03 : 38 : 49.77 & -35 : 23 : 35.6 &  900 &   85 & 20.96 & 1.21 &  \\ 
FCOS 1-2050 & 03 : 39 : 19.06 & -35 : 34 : 07.0 & 1635 &  104 & 20.96 & 1.02 &  \\ 
FCOS 0-2092 & 03 : 39 : 05.02 & -35 : 26 : 53.9 &  970 &  205 & 20.96 & 1.14 &  \\ 
FCOS 0-2065 & 03 : 38 : 21.84 & -35 : 29 : 23.3 & 1804 &  125 & 20.98 & 1.22 &  \\
\hline\hline 
\end{tabular}
\end{table*}
%%%% next line to be removed in 2-column version
\renewcommand{\baselinestretch}{1.0}
\subsection{Radial velocity distribution}
\label{vrad}
\begin{figure}[h!]
\psfig{figure=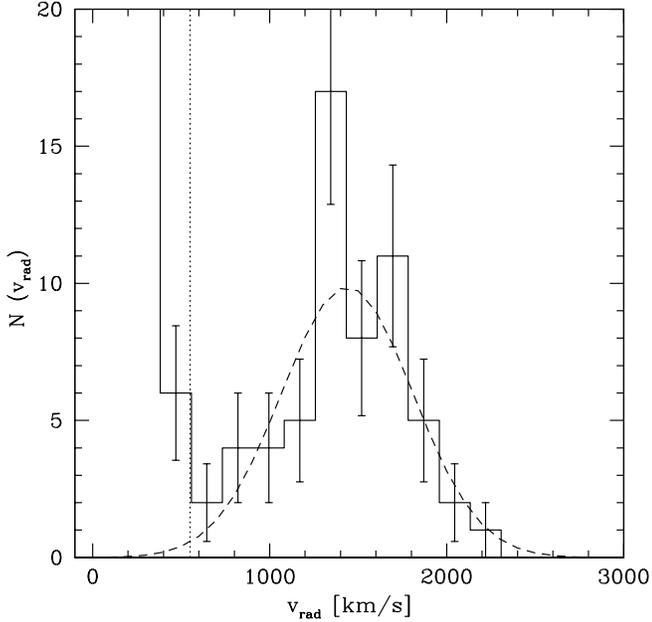,width=8.7cm}
\caption[]{\label{histvrad}Velocity histogram of the objects observed in the FCOS.
The lower limit of 550 km/s applied for Fornax membership is indicated by the dotted line.
The dashed line is a Gaussian fit to the histogram for velocities higher than 550 km/s.
Parameters are:  ${v_{\rm rad,0}}$=1438 $\pm$ 56 km/s, $\sigma_{v}$=372 $\pm$ 46 km/s.}
\end{figure}
\begin{figure}[h!]
\psfig{figure=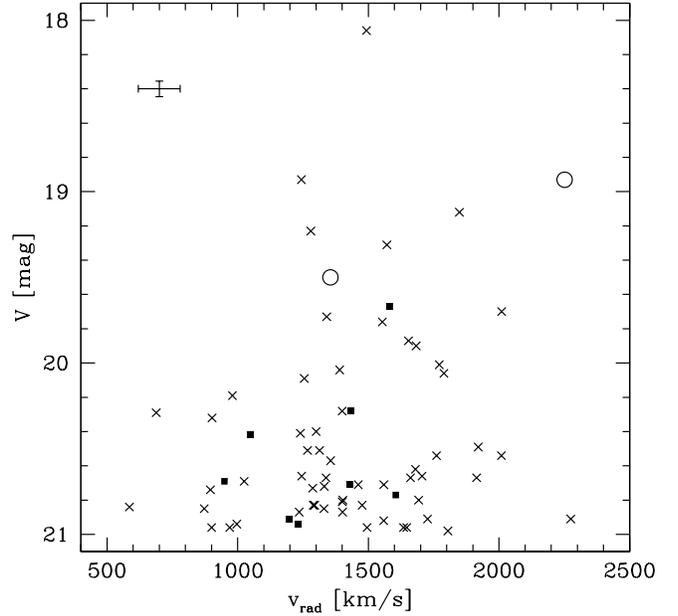,width=8.7cm}
\caption[]{\label{vradv}Apparent $V$ magnitude plotted vs. radial velocity. Crosses: all
FCOS Fornax members. Filled squares: Fornax members
from Dirsch et al. (\cite{Dirsch04}, private communications), not observed in the FCOS. Circles: 
the two UCDs not observed spectroscopically in the FCOS, with their radial velocities
taken from Karick et al. (\cite{Karick03}). The UCDs
are those brighter than V=19.5 mag. Typical errors are indicated by the error bars in the
upper left.}
\end{figure}
\begin{figure}[h!]
\psfig{figure=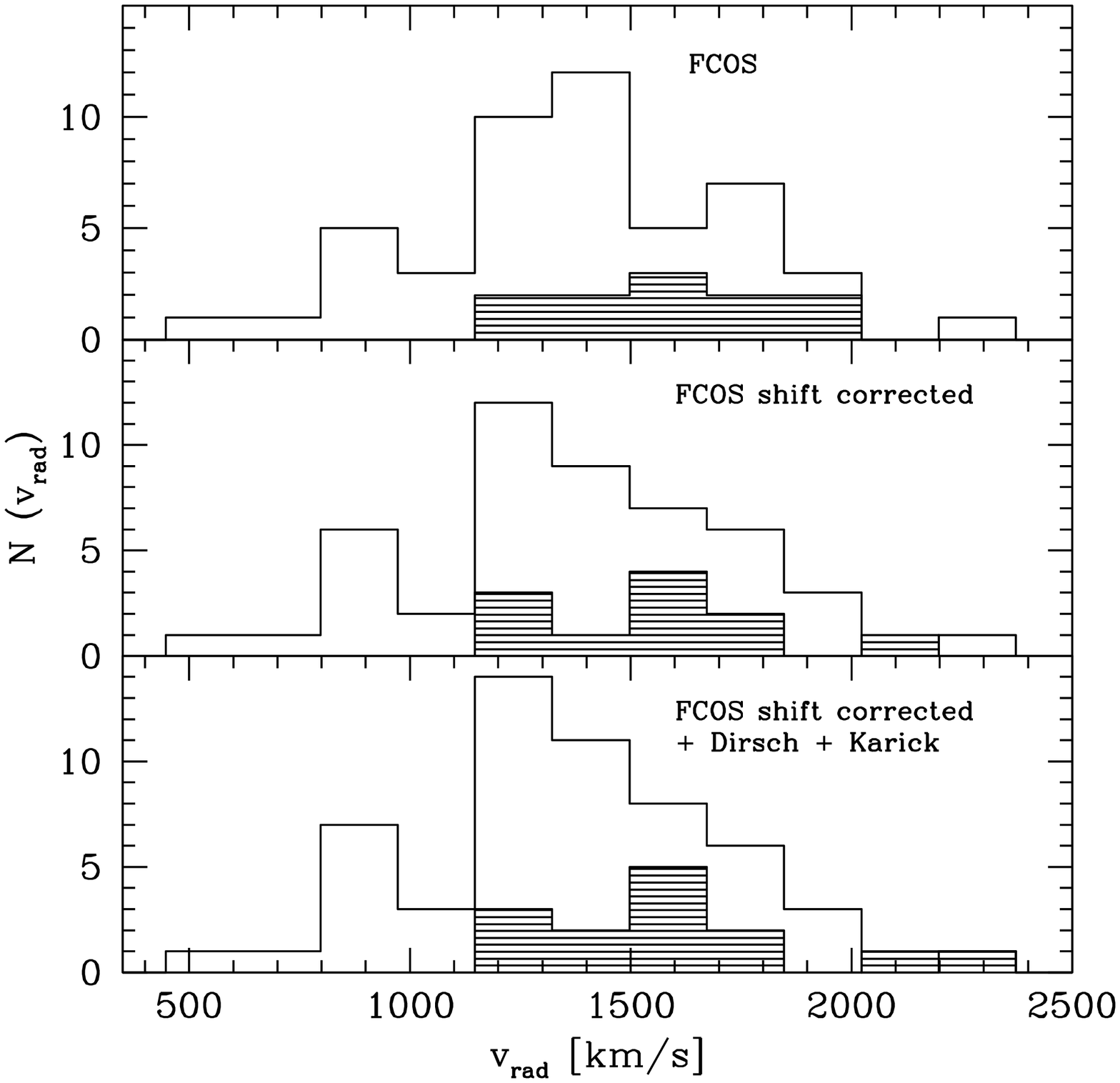,width=8.7cm}
\caption[]{\label{histvrad2}Solid velocity histograms: Fornax members with $20<V<21$ mag. 
Shaded velocity histograms: Fornax members with $V<20$ mag.
{\it Top panel:} FCOS objects. 
{\it Middle panel:} FCOS objects corrected
for the systematic velocity shift between FCOS-1 and FCOS-2 
(see Sects.~\ref{syserr} and ~\ref{vrad}). {\it Bottom panel:} FCOS shift corrected + additional Fornax 
members from Dirsch et al. (\cite{Dirsch04}) + the 2 UCDs not observed in the FCOS, taken from Karick et al.
(\cite{Karick03}).  }
\end{figure}
Fig.~\ref{histvrad} shows
the velocity histogram of the objects observed in FCOS. 
The mean velocity of the Fornax sample
is {1425 $\pm$ 45 km/s, with a velocity dispersion of 326$^{+42}_{-32}$ km/s. Note
that here and in what follows, the velocity dispersion is given by the quadratic difference of
the standard
deviation of the measured velocities around their mean and the mean velocity
error. The standard deviation for
the FCOS sample is 14 km/s higher than the velocity dispersion, with the mean velocity
error being 97 km/s.\\
Applying a
Gaussian fit to the velocity histogram, the parameters are 
${v_{\rm rad,0}}$=1438 $\pm$ 56 km/s, $\sigma_{v}$=372 $\pm$ 46 km/s.
Although the overall velocity histogram for FCOS Fornax members in Fig.~\ref{histvrad}
is symmetric and well
approximated by a Gaussian, Fig.~\ref{vradv} shows that 
objects brighter than $V=20$ mag appear to have a higher mean velocity than the faint ones and a slightly
smaller velocity dispersion.\\
Defining $V=20$ mag as the limit 
between bright and faint Fornax members, a Kolmogorov-Smirnov test shows that the bright and
faint velocity distribution are not drawn from the same underlying distribution at 74\%
confidence. To investigate this in more detail, we calculate the mean velocities and velocity dispersions
for the bright and faint objects separately. These calculations are performed not only for the
entire FCOS-sample, but also for two modified samples:\\
We first correct FCOS-1 and FCOS-2 for the systematic
velocity difference found in Sect.~\ref{vrad} by adding 50 km/s to the FCOS-1 values
and subtracting 35 km/s from the FCOS-2 values. The latter correction serves to remove the 
influence of systematic velocity errors on the velocity distributions, especially when
considering the bright and faint sample separately.\\
Second, we add to the shift corrected FCOS sample 
the eight additional radial velocities 
from Dirsch et al. obtained
for objects not observed in the FCOS and the velocities of the two UCDs not observed in the FCOS (Karick
et al. \cite{Karick03}). This latter sample will hereafter be referred to as shift corrected enlarged sample.\\
The Kolmogorov-Smirnov test yields 74\% disagreement between bright and faint ones
for the shift corrected sample and 88.8\%
disagreement for the shift corrected enlarged sample.
The respective velocity histograms are shown in
Fig.~\ref {histvrad2}.
The results of the calculations are shown in Table~\ref{threesome}.
One can see that the difference in mean velocity between the bright and faint
subsample is over 200 km/s for the shift corrected enlarged sample. This
is significant at the 96\% (2.1 $\sigma$) 
confidence level, as derived from a t-test.
The velocity dispersion is only marginally smaller for the bright objects, 
with the difference disappearing for the shift corrected enlarged sample.\\
We therefore conclude that we have found strong indications for the bright FCOS
Fornax members having a higher mean velocity than the faint ones, while the
velocity dispersions of both sub-samples differ insignificantly. The implications
of these findings with respect to the nature of bright compact objects in Fornax will be discussed
in Sect.~\ref{discussion}, where the results will also be compared with results of other authors.\\
The following two expressions will be used from now on to distinguish between the bright and faint 
sample: ``bright compact
objects'', referring to FCOS Fornax members with $V<20$ mag; and ``faint compact objects'',
referring to FCOS Fornax members with $V>20$ mag. With ``FCOS objects'', we will refer ourselves
to all FCOS Fornax members.\\
\begin{table*}
\begin{tabular}{l||rrrr||rrrr}
Samples&$\overline{v}_{all}$ & $\overline{v}_{bright}$ & $\overline{v}_{faint}$ & Dif. Conf.
&$\sigma _{v,all}$ & $\sigma _{v,bright}$ & $\sigma _{v,faint}$ & Dif. Conf. \\\hline\hline
FCOS & 1425 $\pm$ 45 & 1565 $\pm$ 77 & 1396 $\pm$ 51  & 85 \% & 326$^{+42}_{-32}$
& 236$^{+86}_{-40}$ & 
335$^{+47}_{-36}$& 74\%\\
FCOS shift-corrected (sc) & 1408 $\pm$ 44 & 1541 $\pm$ 83  & 1381 $\pm$ 50 & 83 \% & 322$^{+42}_{-32}$ & 
255$^{+93}_{-43}$ & 328$^{+46}_{-35}$ & 57\%\\
FCOS sc + Dirsch + Karick &  1408 $\pm$ 41 &  1584 $\pm$ 85 & 1367 $\pm$ 45 &  96 \% & 
326$^{+39}_{-30}$ & 302$^{+96}_{-45}$ & 320$^{+42}_{-32}$& 12\%\\
\hline\hline
\end{tabular}
\caption[]{\label{threesome}Table showing the mean velocities $\overline{v}$ and velocity dispersions
 $\sigma_v$ of all, only bright ($V<20$ mag) and only faint ($V>20$ mag) Fornax members. 
Samples considered
are: the entire FCOS-sample (FCOS-1 and FCOS-2); the FCOS-sample corrected
for the systematic velocity differences between FCOS-1 and FCOS-2 (see Sects.~\ref{syserr} 
and ~\ref{vrad}); and the shift corrected FCOS-sample + additional Fornax members
from Dirsch et al. (\cite{Dirsch04}) + the 2 UCDs not observed in the FCOS, taken from Karick et al.
(\cite{Karick03}). 
The column 
``Dif. Conf.'' gives the confidence limit in percent for the hypothesis of the bright and 
faint value being different,
derived with a t-test for the means and with an F-test for the dispersions.}
\end{table*}
\subsection{Magnitude distribution}
\begin{figure}[h!]
\psfig{figure=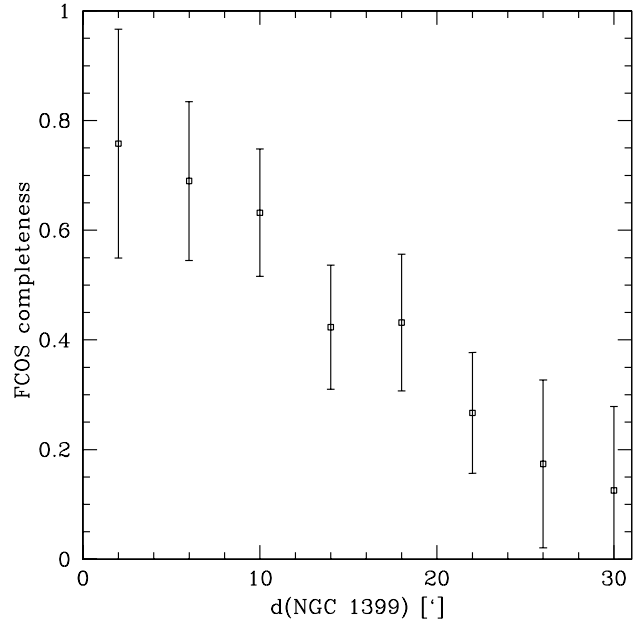,width=8.7cm}
\caption[]{\label{complhist}Completeness of FCOS plotted vs. radial distance from
NGC 1399. The completeness is defined as the ratio of the number of objects with
successfully observed radial velocity and the number of existing candidates. 
For the latter value, both photometric and geometric completeness
has to be calculated. See text for further details.}
\end{figure}
\begin{figure}[h!]
\psfig{figure=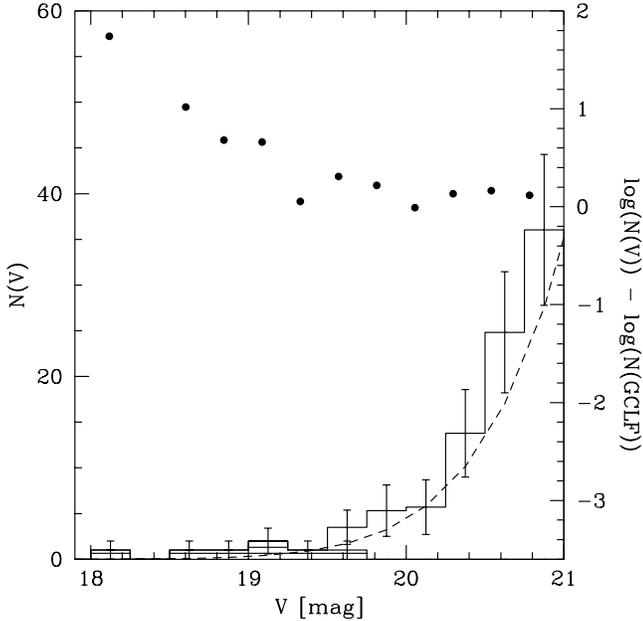,width=8.7cm}
\caption[]{\label{compllf}{\it Solid histogram:} incompleteness corrected luminosity distribution
$N(V)$ of the FCOS objects. Note that for the objects in the central field D1,
only $CR$ photometry is available. Their $R$ magnitudes have been converted
to the $V$-band via $V=R+0.5$ mag. See Sect.~\ref{ctiocand} for further details.
{\it Shaded histogram:} luminosity distribution of the seven UCDs.
{\it Dashed line} Gaussian LF for the GCLF of NGC 1399, taken
from Dirsch et al. (\cite{Dirsch03}). The parameters are: $N_{\rm GC}$=6450, turn over magnitude
V=24.0 mag, width $\sigma$=1.3 mag. {\it Filled circles:} Logarithm of the ratio $\frac{N(V)}{N(GCLF)}$.}
\end{figure}
\noindent In this subsection, we investigate the luminosity distribution of the FCOS objects. 
Can we confirm the soft transition in magnitude space between bright GCs and UCDs from 
{\it paper I}, 
and is the sample sufficiently large to check for an overpopulation at the bright end?\\
\subsubsection{Survey completeness}
\label{surveycompleteness}
In order to investigate the magnitude distribution, 
it is necessary to calculate the overall survey completeness 
$C(r)$, which gives as a function of radius from NGC 1399 
the probability than an object which satisfies our selection criteria is observed successfully. 
\footnote{The magnitude range brighter than $V=19.5$ mag
is excluded from the analysis, as for this range the completeness is given by the FCSS to be 92.4\%
(Drinkwater, private
communications), considered complete for our calculations.}\\
It holds:
\begin{equation}
C(r)=C_{\rm rv}(r) \times C_{\rm det}(r) \times C_{\rm geom}(r)
\end{equation}
$C_{\rm rv}(r)$ is the number of objects with successfully determined radial velocity divided by
the number of photometrically detected candidates. The global, radially integrated $C_{\rm rv}$ is
$\frac{280}{462}=0.606$, see Sect.~\ref{results}. $C_{\rm det}(r)$ is the photometric
detection completeness of candidate objects. $C_{\rm geom}(r)$ is the geometric incompleteness,
caused by incomplete sky coverage of the fields from which candidates where selected. 
$C_{\rm det}(r) \times C_{\rm geom}(r)$ is obtained with artificial star experiments, where
the number density per arcminute of input stars is constant for all investigated images, and the
world coordinates of every input star are calculated. This allows 
to calculate the number density of recovered input stars as a function of distance to NGC 1399,
yielding $C_{\rm det}(r) \times C_{\rm geom}(r)$\\
$C(r)$ is
shown in Fig.~\ref{complhist}. 
In the inner 10$'$, about 70\% of existing
candidates were successfully investigated, about 40\% in the range between 10 and 20$'$. The 
completeness
corrected total number of Fornax members is calculated by dividing the number of detected Fornax
members in each radial bin by the completeness in the same bin. The total completeness $C_0$ for 
detecting
Fornax members then becomes 61.5\%. $C_0$ is corrected for its magnitude dependence
by subdividing the sample into bright ($19.5<V<20.25$ mag) and faint ($20.25<V<21$ mag) objects,
and calculating $C_0$ separately for the two samples. 71\% and 59\% are obtained for bright
and faint objects, respectively. At a given magnitude, $C_0$ is obtained by linear interpolation 
between the two values (see also {\it paperI}).\\
\subsubsection{The form of the luminosity function at the bright end}
\label{lf}
In Fig.~\ref{compllf}, the incompleteness corrected luminosity distribution of FCOS Fornax
members is shown. In addition, the luminosity distribution of the seven UCDs is given.
 The joined luminosity distribution of these two samples proves to
be smooth and without any gaps until $V\simeq 18.5$ mag. This confirms the result derived
in {\it paper I}. We therefore state that there is no separation in magnitude between the UCDs
and the fainter compact objects in Fornax. Overplotted in Fig.~\ref{compllf}
is a Gaussian luminosity function (LF) with the parameters 
$N_{\rm GC}$=6450, turn over magnitude V=24.0 mag and width $\sigma$=1.3 mag, as found
for the GCS of NGC 1399 by Dirsch et al. 
(\cite{Dirsch03}). The paper of Dirsch et al. presents the most complete photometric 
survey of NGC 1399's globular
cluster system, yet. The LF fit to their results
 is a good approximation of the real magnitude distribution
in the magnitude range observed by us, although for very bright magnitudes it slightly underestimates
the number counts. As the real magnitude distribution is smooth and without any significant peaks,
we characterize the shape of its bright end by overplotting on Fig. ~\ref{compllf}
the log-ratio of the number of objects found by us and the number expected from the LF by Dirsch et al. 
 This ratio is close to unity for magnitudes fainter than $V \simeq 20$ mag, 
but rises strongly for brighter magnitudes. 
The magnitude where this ratio change occurs agrees roughly with the limit of 
$V=20$ mag between the objects with higher and lower
mean velocity found in Sect.~\ref{vrad}. For
the entire range $V<20$ mag, the incompleteness corrected overpopulation with respect to the Gaussian luminosity
function from Dirsch et al. is 120\% $\pm$ 65\%, corresponding to 8 $\pm$ 4 objects. This is only slightly
lower than the estimated total number of 14 UCDs as predicted by Bekki et al. (\cite{Bekki03}).\\ 
Note, however, that the overpopulation
found here is with respect to the {\it extrapolation}
of the GCLF fitted at fainter magnitudes, which is not sufficient to prove that there really is one. 
See Sect.~\ref{discussion}
for further discussion.\\
\begin{figure}[h!]
\psfig{figure=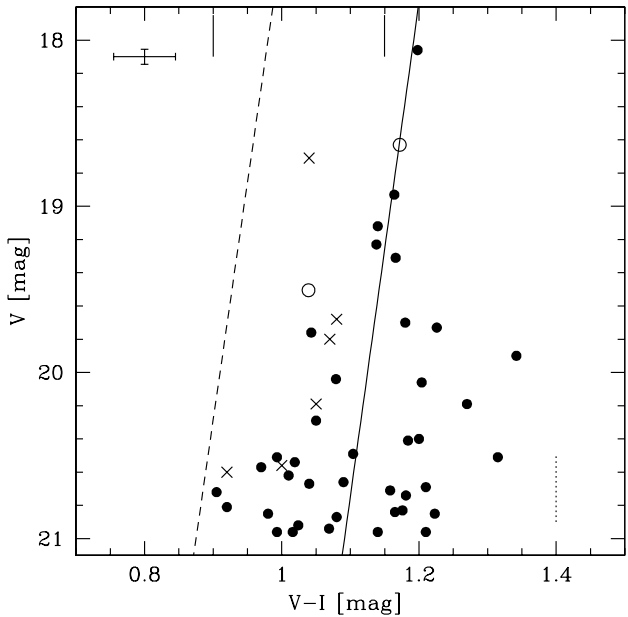,width=8.7cm}
\caption[]{\label{cmd}CMD of the compact Fornax members. Filled circles
are the FCOS objects with $VI$ photometry available (from Hilker et al. \cite{Hilker03}). The
open circles are the 2 UCDs not observed in the FCOS.
Crosses mark $VI$ photometry of Fornax and Virgo dE,N nuclei, taken from Lotz et al. (\cite{Lotz01}).
The solid line is a linear fit to the circles applying a 2$\sigma$ rejection
algorithm. When not applying the 2$\sigma$ rejection, the resulting slope increases by
about 30\% and the overall fit is about 0.02 mag redder than the colours of the brightest
objects.
The dashed line is 
the CM relation
found for Fornax dEs by Hilker et al. (\cite{Hilker03}). The vertical ticks at $(V-I)=$ 0.90 and
1.15 mag indicate the location of the blue and red peak of NGC 1399's bimodal GC 
colour distribution from Gebhardt \& Kissler-Patig (\cite{Gebhar99}). The dotted vertical tick at
$(V-I)=$ 1.40 mag indicates the red colour limit for FCOS-2 candidate selection. For
FCOS-1, the limit had been 1.50 mag (see also Sect.~\ref{lcocand}).}
\end{figure}
\subsection{Colour-magnitude diagram}
\label{cmdtext}
\noindent The mean colour of GC systems in early type galaxies is between $(V-I)=1.00$ and 1.05 
(e.g. Kundu \& Whitmore \cite{Kundu01a} and \cite{Kundu01b}). The blue and red peaks
of the frequently detected bimodal colour distribution generally lie at about 0.95 and 1.18 mag,
respectively (Kundu \& Whitmore \cite{Kundu01a}). No CM relation in the sense that within one globular
cluster system, the GC colour becomes redder with brighter GC magnitude, has been reported for any
GC system. This is in contrast to dEs, for which a tight CM 
relation exists (e.g. Hilker
et al. \cite{Hilker03}).\\
In the case of NGC 1399, Gebhardt \& Kissler-Patig (\cite{Gebhar99}) find from HST WFPC2 archive 
data a bimodal
colour distribution with peaks at $(V-I)\simeq0.9$ and 1.15 mag, respectively. Dirsch et al. 
(\cite{Dirsch03})
find with VLT FORS1-photometry a unimodal colour distribution for bright GCs, with a mean between
1.05 and 1.1 mag.\\
Fig.~\ref{cmd} shows a CM diagram in $V$ and $(V-I)$ of the FCOS objects, 
including the UCDs. $VI$ photometry of dE,N nuclei derived by
Lotz et al. (\cite{Lotz01}) for Fornax and Virgo cluster dE,Ns is shown as well.\\
The CMD has three main features:\\
1. There is a CM relation for the FCOS objects
with a very similar slope to the relation found for Fornax dEs by Hilker et al. (\cite{Hilker03}), 
but shifted
about 0.20 mag redwards. The slope $\frac{\delta(V-I)}{\delta(V)}$ 
found for the FCOS objects is -0.034 $\pm$ 0.011, significant at 3 $\sigma$.
This slope was fitted using a 2 $\sigma$ clipping algorithm. When not doing any clipping,
the slope rises to -0.044 $\pm$ 0.011, but provides a poorer fit to the brightest data points.
For the Fornax dEs the slope is -0.036.
When excluding the brightest 3 data points from
the sigma clipping fit, the resulting slope is -0.038 $\pm$ 0.015.
This shows that the existence of the slope
is not only defined by the brightest data points. When excluding
instead the faint compact objects ($V>20$ mag) from the same fit, the slope becomes -0.041 $\pm$ 0.027, 
i.e. still consistent with the overall slope but only half as significant. Finally,
excluding from the data points used for the last fit also the brightest data point, changes the
slope only slightly to -0.035 $\pm$ 0.039. All these fits to the different restricted data samples indicate
that the existence of the slope stands on firm ground.
For the faint compact objects, the scatter
in $(V-I)$ is about 0.10 mag, too large to reliably measure a slope. The red 
colour limit of $(V-I)=1.40$ mag for
FCOS-candidates has no effect
on the slope of the relation, as the colour space $1.20<(V-I)<1.40$ mag is only very sparsely populated. Note
that already Karick et al. (\cite{Karick03}) have found a colour-offset between dwarf galaxies and UCDs in Fornax,
but their colour precision and sample size was not sufficient to detect a possible CM-relation for the UCDs.\\
2. The colours of dE,N nuclei from Lotz et al. (\cite{Lotz01}) lie between dEs and FCOS objects, 
shifted about 0.1 mag redwards with respect to dEs and about 0.1 mag bluewards of FCOS objects.\\
3. The mean colour of bright compact objects ($(V-I)=1.16 \pm 0.02$) is redder than the average
mean for GC systems, consistent with the metal-rich peak of the bimodal
colour distribution. Their overall colour range is $1.0<(V-I)<1.3$.
The mean colour of the faint compact objects ($(V-I)=1.10 \pm 0.02$) 
and their colour range of $0.9<(V-I)<1.3$ mag is slightly more consistent
with the overall mean colours of GCs.\\
The existence of a CM relation for the bright compact objects as found here is a property that 
has not been reported
for any GCS, yet.
In Sect.~\ref{discussion} this finding will be further discussed.\\
\subsection{Radial density distribution}
\label{raddensdis}
\begin{figure}[h!]
\psfig{figure=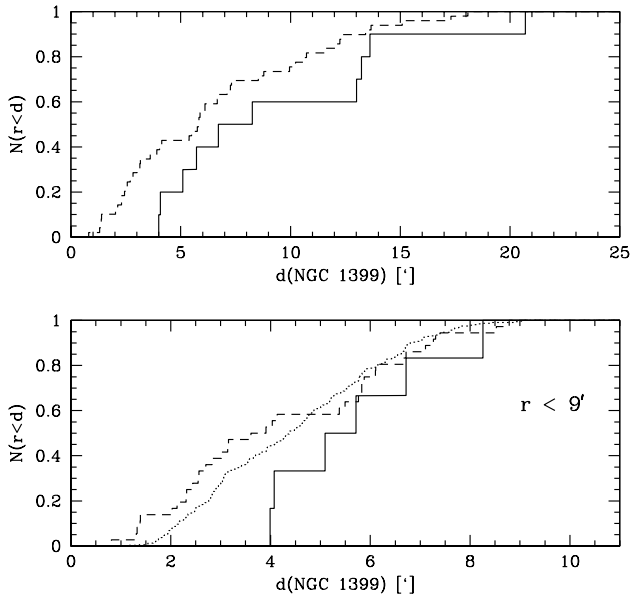,width=8.7cm}
\caption[]{\label{radcum} {\it Top panel:} Cumulative radial distributions of FCOS objects. 
Solid histogram: 
Bright compact objects ($V<20$ mag). Dashed histogram: Faint compact objects ($V>20$ mag).
{\it Bottom panel:}  Comparison between cumulative radial distribution of FCOS objects and 
GCs from Dirsch et al. (\cite{Dirsch04}), restricted to 9$'$, as this is the extension
of the Dirsch sample. Solid and dashed histogram as in top panel.
Dotted histogram: GCs from Dirsch et al. with $(C-R)>1.4$ (see text).}
\end{figure}
\noindent Fig.~\ref{radcum} shows the cumulative radial distribution of bright and faint compact
objects plus
the radial distribution of NGC 1399's GC system as investigated by Dirsch et al. (\cite{Dirsch04}).
Note that Dirsch et al. find that the GCs contained in the blue peak of NGC 1399's bimodal
 colour distribution
have a significantly higher velocity dispersion, mean velocity and more extended
spatial distribution than the GCs in the red peak.
Therefore we compare our values with the Dirsch et al. sample 
restricted to the same colour range as ours, which is $(V-I)>0.9$ (see Fig.~\ref{cmd}), 
corresponding to $(C-R)>1.4$ (Worthey et al. \cite{Worthe94}).
The following conclusions are extracted from Fig.~\ref{radcum}:\\
1. The bright compact objects are slightly less concentrated towards NGC 1399 than 
the faint ones.
Applying a KS-test shows that the sample of bright objects is drawn from a different distribution than the
faint objects at 88\% probability.\\
2. The radial distribution of NGC 1399's GCS within 9$'$ agrees slightly better with the faint than the
bright compact objects.
Applying a KS-test
shows that the faint compact objects are drawn from a different distribution than the 
entire GCS at 71\%
confidence, while the bright ones are inconsistent at the 86\% confidence level.\\
These findings agree with the separation between bright
and faint compact objects in radial velocity (Sect.~\ref{vrad}) and hence add further
credibility to that distinction.\\
\section{Discussion}
\label{discussion}
In the following, we will discuss to what extent our findings provide support for the
hypothesis that the UCDs and possibly other bright stellar clusters around NGC 1399
do not belong to NGC 1399's GCS but rather are remnants of threshed
dE,Ns. Also, we consider the possibility of stellar super-clusters created in past merging
events.\\
\subsection{Explanation for dynamically distinct subgroups}
\label{disdyn}
A difference in mean radial velocity has been found between the bright ($V<20$ mag) and 
faint ($V>20$ mag) compact objects
in Sect.~\ref{vrad}. See Table~\ref{threesome} for exact numbers.
We compare these results with values published in the literature for the 
Fornax cluster.\\
Dirsch et al. (\cite{Dirsch04}, and private communications) obtain a mean velocity of 
1440 $\pm$ 15 km/s with a dispersion of 325 km/s $\pm$ 11 km/s for a sample of almost 500 GCs brighter than
$V=23$ mag, when restricting their sample to radial velocities higher than 550 km/s, as done
in this paper. These values are in agreement with our results for the entire FCOS sample, 
see Table~\ref{threesome}. The same table shows that the mean velocity of the Dirsch sample is
slightly more consistent with the faint than the bright compact objects. 
Due to the different spatial and velocity distribution found by Dirsch et al. for the red and blue
GC sample, it seems however more appropriate to compare our velocity values with their colour restricted
sample, 
i.e. for $(C-R)>1.4$, as already pointed out in see Sect.~\ref{raddensdis}.\\
The mean velocity of Dirsch et al.'s colour restricted sample 
is 1422 $\pm$ 17 km/s with a dispersion of 309 $\pm$ 13 km/s.
The velocity dispersion agrees well with our entire sample and the sub-samples. A t-test shows
that the mean velocity of Dirsch's sample is different at 79\% confidence (1.2 $\sigma$) from the {\it faint} compact objects 
in the shift corrected enlarged sample and different at 93.5\% confidence (1.9 $\sigma$) from the {\it bright} ones
in the same sample. For the shift-corrected not enlarged sample the confidence limits for different means 
are 60\% for the faint objects and 77\% for the bright ones.\\
Thus, our findings show that the mean velocity of faint compact objects is more consistent with
NGC 1399's GCS than that of the bright ones.\\
In that context it is natural to ask how our values, especially for the bright compact objects, 
compare with what is found for the dwarf galaxy 
population of Fornax, because the threshing scenario predicts dwarf galaxies as the progenitor objects
of UCDs.\\
The most recent and complete spectroscopic observations 
of Fornax cluster galaxies come from Drinkwater et al. (\cite{Drinkw01a} and \cite{Drinkw01b}).
Separating their Fornax main cluster sample 
into dwarfs and giants with $b_j=15$ mag as the limiting magnitude, 
one obtains a mean radial velocity of 1511 $\pm$ 56 km/s and a dispersion of 406 $\pm$ 36 km/s 
for the
dwarf galaxies, while the giants have 1426 $\pm$ 52 km/s and 314 $\pm$ 33 km/s. The 
latter values are consistent with the faint compact objects and Dirsch et al.'s GCs.
Drinkwater et al. suggest that the difference in dispersion is due to the fact that 
the Fornax dwarf galaxies are not yet virialized, but still falling into the cluster. 
This might as well explain the small offset in radial velocity between dwarfs and giants, assuming
that the dwarfs have an anisotropic infall.\\
Comparing the different mean velocities for Fornax dwarf and giant galaxies from Drinkwater et al.
with our results using a t-test shows that the mean velocity of the bright compact objects is 
consistent with that of the dwarf galaxies, yielding only 45\% confidence for the hypothesis of
a different mean. It is inconsistent at 95\% confidence with that of the giants. For the comparison of faint
compact objects to dwarfs and giants, the confidence levels for a different mean are 90\% and 60\%, 
respectively.\\
This supports the hypothesis that UCDs and bright compact objects ($V<20$ mag) originate mainly 
from dwarf galaxies, while the faint compact objects ($V>20$ mag) are mainly genuine GCs.\\
However, there is a difference of 104 $\pm$ 102 km/s between the velocity dispersions of 
bright compact objects in the shift corrected enlarged sample and Fornax dwarf
galaxies. Although this is only marginally significant, a possible reason for this difference
is given by Bekki et al. (\cite{Bekki01}, \cite{Bekki03} and private
communications). They find from their simulations that both highly eccentric orbits ($e_p > 0.7$) and smaller 
pericenter distance are required for the transformation from dE,Ns to UCDs. These restrictions mean that
the UCDs have a smaller velocity dispersion than the non-threshed dEs or dE,Ns. Thus the
dispersion of UCDs around the mean cluster velocity 
should be smaller than that of average dEs or dE,Ns, which qualitatively explains our findings.
Note that the dispersion of the bright compact objects around the mean velocity of the 
Fornax giants (1426 km/s) is 335 km/s for the shift corrected enlarged sample, i.e. still somewhat
smaller than that of average dwarf galaxies.\\
An interesting, speculative consideration is: if the threshing affects a dE population with smaller
velocity dispersion than the mean cluster dispersion, then the velocity standard deviation of this
population is lower and therefore the velocity values are concentrated
stronger around the mean than for the rest of the population. This means that by excluding
the threshed population, which has a smaller velocity dispersion, from the entire sample of dEs, 
the velocity distribution of
the remaining non-threshed dEs should broaden. 
To estimate the magnitude of this effect, we add to the 
velocity distribution of dwarf galaxies from
Drinkwater et al. (\cite{Drinkw01a} and \cite{Drinkw01b}) {\it twice} the velocities of bright compact objects
from this paper in order to also take into account entirely disrupted dEs.
The resulting velocity dispersion is 370 km/s, compared to 406 km/s for
the original sample. This speculative calculation shows that the threshing might be 
partially responsible for the higher velocity dispersion of Fornax dwarf galaxies. Further discussion of
this is, however, beyond the scope of this paper.\\
An additional process to create UCDs has been brought forward by Fellhauer \& Kroupa (\cite{Fellha02}), 
who
show that massive stellar super-clusters formed in violent merging events can after few Gyrs resemble
the properties of UCDs. In the case of the very dense Fornax cluster, the probability of a major merging
occurred in the past is very high. The GCS of NGC 1399 has a pronounced bimodality (Gebhardt \&
Kissler-Patig \cite{Gebhar99}, Dirsch et al. \cite{Dirsch03}), suggesting that at least two separated star
forming events have occurred. If the bright compact objects were such super-clusters, 
one might
expect to find a trace of the merging event like a different velocity than the Fornax mean,
alignment along a preferred axis or association with major Fornax galaxies.
While a different (higher) velocity has been found, Fig.~\ref{mapobs2} shows 
that there is no 
alignment for the bright compact objects, 
nor are they distributed especially close to a Fornax cluster galaxy
other than the central galaxy NGC 1399. However, this does not rule out the super-cluster 
scenario,
as after several Gyrs the traces of the merger event, be it in velocity or 2D-space, become smeared out.
In fact, detailled simulations regarding this scenario for the Fornax cluster case would be needed to better assess
its validity.\\
\subsection{Colour magnitude relation for the FCOS Fornax members}
It has been shown in Sect.~\ref{cmdtext} that there is a CM relation for the FCOS Fornax
members with a slope agreeing with what is found for dEs. This relation is shifted about 
0.2 mag redwards with respect to the dE relation. The colours of dE,N nuclei lie between those of
dEs and FCOS objects. While the bright compact objects ($V<20$ mag) follow a slope consistent with the overall
one, the faint compact objects ($V>20$ mag) scatter too much in $(V-I)$ in order to reliably confirm
 or discard the existence
of a slope in that magnitude range.
With a Gedankenexperiment it can be estimated, where
the CM relation for threshed nuclei should be situated with respect to that of the dEs:\\
One assumes that the nucleus has the same colour as the entire dE,N and that
in the course of the threshing, the stellar population of a dE,N and especially that of
its nucleus is not
altered. 
Hence, only the dE,N mass and thereby its luminosity are decreased in the threshing process until the naked nucleus is 
left over and an UCD is created. In a CMD this means that in the course of
the transformation from dE,N to UCD, a galaxy moves in the y-axis (magnitude) by the amount
corresponding to its luminosity-loss. The CM relation defined by the newly formed UCDs would then be shifted
redwards by its slope multiplied with the magnitude difference between dE,N and UCD, assuming that this difference
does not depend strongly on the dE,N magnitude.
Bekki et al. (\cite{Bekki01} and \cite{Bekki03}) calculate for their
fiducial model that in the
threshing process, the magnitude difference between the original dE,N and the final UCD is 4.1 mag. Such 
a magnitude difference 
corresponds to a redward shift of the CM relation of about 0.15 mag, as the slope of the dE CM relation
is about -0.036. This brings the shifted CM relation of dEs very close (about 0.07 mag) to the 
one found in this paper for the FCOS objects. The data of Lotz et al. (\cite{Lotz01}) 
suggest that the mean magnitude
difference between a dE,N and its nucleus is about 5 mag, 1 mag larger than the difference given
by Bekki et al. for their fiducial model. In that case, the shift of the CM relation between dE,Ns and 
stripped nuclei would be about 0.18 mag 
and bring to match both relations within the colour scatter of our data.\\
Thus, with a simple but plausible assumption the threshing scenario predicts 
a CM relation for threshed nuclei with the same slope and the same zero point (to within 0.05 mag) as 
observed for the FCOS objects.\\
However, the condition that the dE,N nuclei have the same colour as the entire galaxy,
is probably an oversimplification. HST photometry of Virgo and Fornax dEs and dE,N nuclei 
(Stiavelli et al. \cite{Stiave01} and Lotz et al. \cite{Lotz01}) shows that
the nuclei are on average about 0.07 mag bluer than the underlying galaxy light. 
If the progenitor dE,Ns of UCDs would have the same properties as the dE,Ns and nuclei investigated
by Lotz et al., the UCDs would be expected to be about 0.10 mag bluer than they actually are. 
Indeed, the nuclei of Lotz et al.
are between 0.05 and 0.10 mag bluer than the bright compact objects, including the UCDs (see Fig.~\ref{cmd}).\\
This discrepancy could be explained, if the already stripped nuclei tended to contain older
populations than 
the dE,Ns and their nuclei as investigated by Lotz et al. (\cite{Lotz01}).
This is a plausible assumption for two reasons:\\
(1) The tidal threshing strongly reduces the gravitational potential well of the threshed galaxy both
in its center and its outskirts, and thus
less gas than compared to non-threshed dE,Ns can be retained for possible star forming bursts.\\
(2) About 5 Gyrs is a minimum age for UCDs (Bekki et al.
\cite{Bekki03}), as this is about the time-frame needed for complete stripping.\\
To get an idea of the age differences needed to explain the observed colour difference between UCDs and
dE,N nuclei,
stellar population models by Worthey (\cite{Worthe94}) are used. We find that for a population of Fe/H=-0.225 and
age 5 Gyrs, $(V-I)=1.08$. For the same metallicity but age 10 Gyrs, we obtain $(V-I)=1.18$. In other words, an
age difference of the time needed to transform a dE,N into a UCD can explain the colour difference
between the bright compact objects and
the dE,N nuclei.\\
An alternative explanation for our finding could be a globular cluster CM relation in the sense 
that within one GCS, brighter GCs become increasingly redder. To our knowledge
the relation found by us would then be the first one reported for a GCS, yet.
The idea is that brighter and more massive clusters, e.g. the stellar super-clusters created in mergers
(Fellhauer \& Kroupa (\cite{Fellha02}), become increasingly more 
self-enriched with growing mass 
(e.g. Platais et al. \cite{Platai03}, Meylan et al. \cite{Meylan01}). Very massive GCs
would on average have redder colours than less massive ones (of the same GCS), just as found by us. 
The merger event could also be responsible for the dynamical differences found by us between bright and
faint compact objects.
However, the idea of a mass-metallicity relation has 
been questioned
by Parmentier \& Gilmore (\cite{Parmen01}), who have shown that the old galactic halo globular
clusters seem to follow a reverse magnitude-metallicity relation, with less massive clusters having 
higher
metallicities. They claim that this anti-correlation is as predicted by self-enrichment models.\\
Note that a CM relation for massive GCs as it might have been found by us,
cannot be backed up with earlier findings of such a relation but stands on its own.
In contrast, the CM relation for dwarf galaxies, from which the UCDs originate according
to the threshing scenario, is a well established fact (Hilker et al. \cite{Hilker03}).\\
We therefore believe that based on our present knowledge, the CM relation found by us is more readily 
explained with the threshing scenario and 
a galactic origin of the UCDs rather
than with massive GCs following a CM relation. 
The fact that the slope is well defined for $V<20$ mag is consistent with the proposed magnitude 
separation of 
$V=20$ mag between threshed nuclei and normal GCs, as proposed in the previous section. There can 
also be fainter
UCDs with $V>20$ mag or GCs with $V<20$ mag, but our findings suggest 
that in the respective magnitude regimes they are in the minority.\\
\subsection{Overpopulation at the bright end}
In Sect.~\ref{lf} it was shown that the number of bright compact objects is 
more than two times higher
than expected from the GCLF of Dirsch et al. (\cite{Dirsch03}), 
which is the best determined
GCLF of NGC 1399 up to now. For $V<20$ mag, there are 8 $\pm$ 4 more objects than expected
from integrating the GCLF of Dirsch et al. This overpopulation is 
consistent with the threshing scenario
as a source of bright compact objects which add to the LF of very bright GCs.
Bekki et al. (\cite{Bekki03}) have shown that the total number of UCDs in the Fornax cluster should 
be about 10-15, which agrees roughly with our finding.
Apart from threshed dE,Ns, it is clear that also the super-clusters proposed by 
Fellhauer \& Kroupa (\cite{Fellha02})
could add to the number counts at bright magnitudes.\\
However, the overpopulation found by us is only 
with respect to the {\it extrapolation} of the LF fit by Dirsch et al. to their
data at fainter magnitudes (the turn-over magnitude is at about $V=24.0$ mag). The contamination
by background galaxies or foreground stars for $V<20$ in the FCOS is about 90\%, illustrating
that Dirsch et al. could not use this part of the magnitude distribution
for fitting their LF.\\
Therefore, {\it only if} the entire GCLF of NGC 1399 would be described well by Dirsch et al.'s
fit also at brighter magnitudes, the luminosity distribution of FCOS members found by us 
supports a slight overpopulation for $V<20$ mag, caused either by threshed dE,Ns or
stellar super-clusters.\\
\section{Summary and conclusions}
\label{summary}
In this paper, we have presented the second part FCOS-2 of the Fornax Compact Object
Survey FCOS and investigated the merged data sets of FCOS-1 ({\it paper I}) and FCOS-2.
The aim of FCOS was to obtain more clues on the nature of the so-called
``Ultra Compact Dwarf Galaxies'' (UCDs), as detected in the central Fornax cluster and described 
by Drinkwater et al. (\cite{Drinkw03}). 280 objects in the magnitude
space $18<V<21$ mag, covering both UCDs and bright globular clusters in Fornax, 
were observed spectroscopically to determine
their cluster membership. Their distribution in radial velocity, colour,
magnitude and space was investigated. The following results have been obtained:\\
1. 54 new compact Fornax members in the magnitude range $19.7<V<21$ mag 
were detected. In addition, 5 of the 7 previously discovered UCDs were observed,
covering a magnitude range of $18<V<19.5$ mag.\\
2. The radial velocity distribution is 
 different for bright ($V<20$ mag) and faint ($V>20$ mag) compact objects. 
At the 96\% (2.1 $\sigma$) confidence level, the bright ones have a 
higher mean velocity than the faint ones.
The result for the faint compact objects is more consistent with existing studies on the GCS
of NGC 1399 than for the bright ones. The bright compact objects 
have a mean velocity consistent with
the mean dwarf galaxy velocity in the Fornax cluster, but a lower velocity dispersion
than the dwarf galaxies. This smaller velocity dispersion is explained
qualitatively  within the threshing scenario by Bekki et al. (\cite{Bekki03}).
 Our result is therefore consistent with the threshing scenario as a source of bright compact
objects and suggests
that UCDs created by threshing dE,Ns may extend down to $V\simeq$ 20 mag. Our findings are also 
consistent with the super-cluster scenario as a source of UCDs (Fellhauer \& Kroupa \cite{Fellha02}), 
but the range
of observed properties consistent with this scenario is larger than in the threshing scenario.\\
3. The FCOS Fornax members follow a CM relation with a very similar slope
to that of dEs, but shifted about 0.2 mag redwards. This shift is explained well 
by the threshing scenario, assuming that the threshed
dE,N keeps its colour in the course of the tidal interaction and only loses luminosity. 
The bluer colours of dE,N nuclei compared to bright FCOS objects are explained
by assuming an age difference between the two populations of the time needed to transform
a dE,N into a UCD. An alternative
explanation for the CM relation is that the bright FCOS objects are super-massive
GCs created in a merger which follow a mass-metallicity relation in the sense
that mass increases with metallicity. This would be the first time that such 
a relation
is detected for a GCS. \\
4. The magnitude distribution of FCOS Fornax members shows a fluent transition between
UCDs and GCs and a slight overpopulation of 8 $\pm$ 4 objects for $V<20$ mag with 
respect to the extrapolation
of the GCLF by Dirsch et al. (\cite{Dirsch03}),  which
matches roughly the predicted
number of UCDs in the threshing scenario.
This finding is also consistent with the super-cluster-scenario as a source of additional
compact objects in the given magnitude regime. However, it is not sufficient to solidly support either
scenario because
the reference GCLF was determined at fainter magnitudes than investigated here.\\
5. The bright FCOS objects have a more extended spatial distribution than the faint ones at
88\% confidence and have a distribution inconsistent with the GCS of NGC 1399 at 86\% confidence.
The distribution of faint FCOS objects is more consistent with NGC 1399's GCS.\\\\
Our results are consistent with a substantial fraction of 
bright compact objects ($V<20$ mag, $M_V<-11.4$ mag)
being nuclei of threshed dE,Ns, whereas the fainter
ones are mainly normal GCs of NGC 1399's very rich GCS. 
The existence of stellar super-clusters following a CM relation
as an alternative source of bright compact objects 
is also consistent with our data.\\
Defining UCDs as extremely bright star clusters distinct
to genuine globular clusters, our findings show that in Fornax the UCDs extend 
to at least $M_V= -11.4$ ($V=$ 20) mag,
about 0.5 mag deeper as the 7 UCDs found by Drinkwater et al. (\cite{Drinkw00a}).
The UCDs populate the colour range $1.0<(V-I)<1.3$.
For $M_V>-11.4$ ($V>20$) mag, there can also be UCDs with possibly bluer colours than given above
which mix up with genuine GCs.\\
Doing a survey similar to FCOS in Virgo or other massive nearby clusters where higher numbers of UCDs
are expected (Bekki et al. \cite{Bekki03}) would be the next step to compare properties of UCDs 
in different environments. 
\acknowledgements
We thank the referee Dr. Michael Drinkwater for his comments which improved the paper.
The authors are very grateful to Dr. Boris Dirsch for providing us with his data and for helpful
discussions. We thank Dr. Kenji Bekki for providing us with information about the dynamical implications
of the threshing scenario. SM was supported by DAAD Ph.D. grant Kennziffer D/01/35298 and
DFG Projekt Nr. HI 855/1-1. LI acknowledges support from proyecto FONDAP \# 15010003.\\

\enddocument